\documentclass[fleqn,12pt,twoside]{article}
\usepackage{espcrc1}
\usepackage{times}
\usepackage{graphicx}
\usepackage[figuresright]{rotating}

\title{Masses of excited baryons from chirally improved quenched lattice 
QCD\thanks{Presented 
at Baryon 2004 by C.B.\ Lang. 
The work was supported by Fonds zur F\"orderung der 
Wissenschaftlichen Forschung in \"Osterreich (P16824-N08 and
P16310-N08) and by DFG and BMBF.}}

\author{{Tommy Burch$^a$, 
Christof Gattringer$^{a}$, 
Leonid Ya.\ Glozman$^b$, Christian Hagen$^a$, Dieter Hierl$^a$,
Reinhard Kleindl$^b$, C.\ B.\ Lang$^b$, 
and Andreas Sch\"afer$^a$
\hspace{1mm}\\(BGR [Bern-Graz-Regensburg] Collaboration)} 
\vskip5mm
$^a$ Institut f{\"u}r Theoretische Physik, Universit{\"a}t
Regensburg, D-93040 Regensburg, Germany. 
\vskip2mm
$^b$ Institut f{\"u}r Physik / Theoretische Physik, Universit{\"a}t
Graz, A-8010 Graz, Austria.
\vskip3mm}
\begin{document}
\maketitle
\begin{abstract}  
Whereas ground state spectroscopy for quenched QCD is well understood, it is
still a challenge to obtain results for excited hadron states. In our study we
present results from a new approach for determining spatially optimized
operators for lattice spectroscopy of excited hadrons. In order to be able to
approach physical quark masses we work with the chirally improved Dirac
operator, i.e.,  approximate Ginsparg-Wilson fermions. Since these are 
computationally expensive we restrict ourselves to a few quark sources. We use Jacobi 
smeared quark sources with different widths and combine them  to  construct
hadron operators with different spatial wave functions. This allows us to
identify the Roper state and other excited baryons, also in the strange sector.
\end{abstract}

\section{Quark sources and interpolating fields}

Ground state spectroscopy on the lattice is by now a well understood
physical problem with impressive agreement with experiment.
The lattice study of excited states is not as advanced. 
The Euclidean correlation function of an 
interpolating operator contains contributions from all states 
with the correct quantum numbers and
the masses
of the excited states show up only in the sub-leading exponentials. A
direct fit to a sum of exponentials is cumbersome since the
signal is  strongly dominated  by the ground state. Also with 
methods such as
constrained fits, the maximum entropy method \cite{chenetal,sasaki}
and a recently proposed method \cite{guadagnoli}
one still needs high statistics for reliable results.  

An alternative method is the variational method
\cite{variation} where one diagonalizes a matrix containing all 
cross-correlations of a set of several basis operators $O_i$
with the correct quantum numbers,
\begin{equation} 
C(t)_{ij}=
\langle O_i(t)\,O_j^\dagger(0) \rangle 
=\sum_n \;\langle 0|O_i|n\rangle e^{-t\,M_n} \langle n|O_j^\dagger|0\rangle \;,
\end{equation}
(written here for infinite temporal extent of the lattice and operators 
projected to vanishing spatial momentum).
For a large enough and properly chosen set of basis operators each eigenmode
is then dominated by a different physical state. 
Finding these states is equivalent to solving the 
generalized eigenvalue problem \cite{variation} with eigenvalues behaving as
\begin{equation}\label{expdecay}
\lambda^{(k)}(t) \propto  e^{-t\,M_k} 
\left[1+{\cal O}(e^{- t\Delta M_k})\right]\;.
\end{equation}
Each eigenvalue corresponds to a different energy level $M_k$ dominating its 
exponential decay; $\Delta M_k$ is the distance to nearby energy levels.  The
optimal operators  which have maximal overlap with the physical states are
linear combinations of the basis operators. After normalization the largest
eigenvalue gives the correlator of the ground state, the second-largest
eigenvalue corresponds to the first excited state, and so on.

Although the set of basis operators can be made arbitrarily large, in  realistic
calculations the intrinsic statistical errors make the  diagonalization
increasingly unstable.  The challenge is then to provide a set of operators
large enough to span the physically relevant space but still small enough to allow
significant results for the given statistics.

In earlier work  \cite{broemmeletal} we used three different interpolating
fields for the nucleon sector, but could not resolve a signal for the Roper
excitation. Here we concentrate on the first two of those,
\begin{equation}\label{chiops}
\chi_1(x) =
\epsilon_{abc} \left[u_a^T(x)\,C \,\gamma_5\,d_b(x)\right] u_c(x) \;,\quad
\chi_2(x) =
\epsilon_{abc} \left[u_a^T(x)\,C\, d_b(x)\right] \gamma_5 \,u_c(x) \;.
\end{equation}

We concluded that in order to get a reliable Roper signal it is also important
to optimize the  spatial properties of the interpolating operators. It can be
argued that a node in the radial wave function is necessary to  capture reliably
the Roper state or other radially excited hadrons. In \cite{prd-paper} we
demonstrated that an elegant solution is to combine Jacobi smeared quark sources
with different widths to build the hadron operators and compute the
cross-correlations in the variational method. We find good effective  mass
plateaus for the first and partly the second and even the third radially excited states. The
eigenvalues $\lambda^{(k)}(t)$ are then fitted  using standard techniques. 

Already earlier \cite{jacobi1} Jacobi smeared sources were combined with point
sources and cross-correlations studied in similar spirit (see also
\cite{burch}). The technique of Jacobi smearing is well known
\cite{jacobi1,jacobi2}. The  smeared source lives in the timeslice  $t = 0$  and
is constructed by iterated multiplication with a smearing operator  $H$ on a
point-like source. The operator $H$ is the spatial hopping part of the  Wilson
term at timeslice 0; it  is trivial in Dirac space and acts only on the color
indices. This construction has two  free parameters: The number of smearing
steps $N$ and the hopping parameter $\kappa$. These can be used to adjust the
profile of the source. We worked with two different sources, a narrow source
$n$ and a wide source $w$ with parameters chosen such that the profiles approximate Gaussian
distributions \cite{prd-paper}.
The distribution of the two source-types correspond to
half-widths of  $\sigma \approx 0.27$~fm and  $\sigma \approx 0.41$~fm. 

The two sources allow the system to build  up radial wave functions with and
without a node. The parameters were chosen such that simple linear combinations
$c_n \, n \, + \, c_w \, w$  of the narrow and wide profile can approximate the
first and second radial wave functions of the spherical harmonic oscillator: The
coefficients $c_n \sim 0.6, c_w \sim 0.4$ approximate a  Gaussian with a
half-width of $\sigma \sim 0.33$ fm, while $c_n \sim 2.2, c_w \sim -1.2$
approximate the corresponding excited wave function with one node. Note,
however, that the final contribution of the corresponding source terms
is not pre-determined but results from the diagonalization.

Following the variational method \cite{variation} we compute the complete
correlation matrix of operators. The hadron sources we use for the correlation
matrix are constructed from the  narrow and wide quark sources. Respecting
symmetries we then have for both types of nucleon operators (\ref{chiops}) six
combinations  of wide or narrow quark sources, in total 12 interpolating
field  operators and a $12\times 12$ correlation matrix. The final form of the
wave function  is determined through the eigenvectors resulting from the
variational method.

\section{Baryon results}

We use the chirally improved Dirac operator \cite{chirimp} and work in the
quenched limit, ie., without dynamical fermions.  This operator is an
approximate Ginsparg-Wilson operator and was well tested in quenched ground
state spectroscopy \cite{bgr}  where pion masses down to 250 MeV have been
reached at a considerably smaller  numerical cost than needed for exact 
Ginsparg-Wilson fermions.  The gauge configurations were generated on a  $12^3
\times 24$ lattice with the L\"uscher-Weisz action \cite{Luweact} (further
studies for larger lattices are in progress). The inverse gauge coupling is
$\beta = 7.9$, giving rise to a lattice spacing of $a = 0.148(2)$~fm as
determined from the Sommer parameter \cite{scale}. The statistics of our
ensemble is 100 configurations. For the $u$, $d$ quarks we use degenerate quark
masses $m$ ranging from  $a\,m = 0.02$ to  $a\,m = 0.20$. More details are
discussed in \cite{prd-paper}.

\begin{figure}[t]
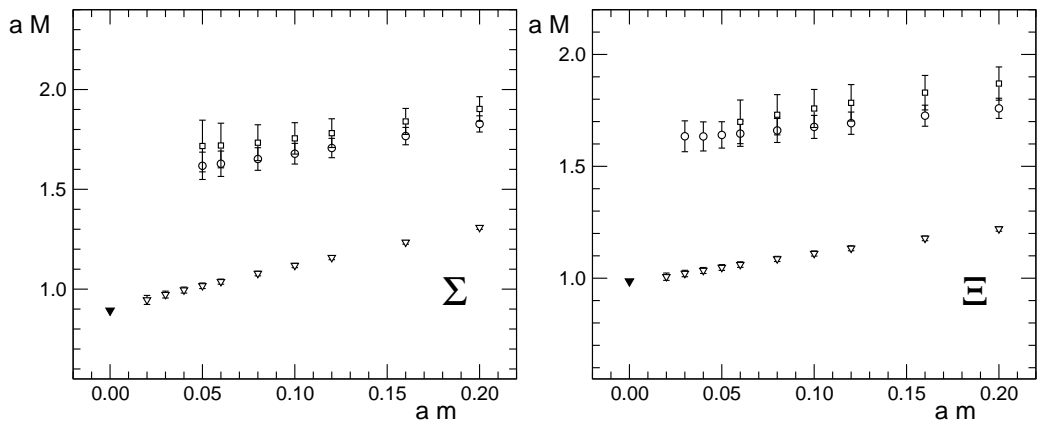

\begin{center}
\includegraphics*[width=6.8cm]{sigma.eps}
\includegraphics*[width=6.8cm]{xi.eps}
\end{center}
\vspace*{-7mm}
\caption{Masses of $\Sigma$ and $\Xi$ baryons in units of the lattice  spacing
$a\approx 0.148$~fm (=1/1.33~GeV) (triangles: ground state, circles: negative
parity ground state, squares:  first positive parity excited state, full
symbol:  experimental value of ground state); the mass of the strange quark has
been  adjusted to give physical $K$-meson mass.
\label{fig1}}
\vspace*{-5mm}
\end{figure}

Although we determine the full correlation matrix as discussed,  it turns out to
be numerically advantageous to analyze only a subset of operators.  The exponential
decay of the three leading eigenvalues is clearly identified. We identify these
signals with the nucleon, the Roper state and  the next positive parity
resonance $N(1710)$.  The larger eigenvalues have too large statistical errors
for a reliable interpretation (see also \cite{chenetal} concerning the problem
of nucleon-$\eta^\prime$ ghost  contributions).

In principle both interpolators $\chi_1$ and $\chi_2$ should couple to all
states with the corresponding quantum numbers; however, the coupling amplitude
depends on the internal structure of the physical states (with possibly
different overlap). Since we can continuously increase the quark mass
and thus make contact with the heavy quark region we also may argue in
such a framework. All Roper states form an excited 56-plet of SU(6) and in this
multiplet the parity of any two-quark subsystem is positive.  The $N(1710)$ 
belongs to another multiplet containing both positive and negative parity
two-quark subsystems. The two-quark subsystem in the brackets in  $\chi_1$ has
positive parity, while the subsystem in the brackets of $\chi_2$ has negative
parity. Hence in $\chi_2$ we should see only the signal from the  $N(1710)$
state, and no signals from the nucleon and the Roper. This is  exactly what we
observe in \cite{prd-paper}, where the relevant figures for the nucleon channel are
shown.

We may also combine quarks with different masses, allowing for a strange quark.
We  fix the strange quark mass such that the $K$-meson mass has its experimental
value  in the chiral limit of the $u,\,d$-masses. We then can determine also the
$\Sigma$- and $\Xi$-baryons using interpolator $\chi_1$ and again different width
quarks sources. First results are exhibited in Fig. \ref{fig1}. We have to
stress, however, that the results given here are for relatively small lattices  (spatial
lattice size 1.8 fm).

A crucial test of our method is to check whether indeed the ground state is
built from a nodeless combination of our  sources and the excited states do show
nodes.  This question can be addressed by analyzing the eigenvectors of the 
correlation matrix. This has been done in \cite{prd-paper} and indeed confirms
the expectation.

\section{Conclusion}

Ground state masses approach, in naive chiral extrapolation, their experimental
values well. The excited state masses are above their experimental
values. Since our results are for small spatial lattice size a volume dependence
study still has to confirm this behavior. Excited states may be more affected
by finite volume effects, but also be more sensitive to the 
quenched approximation. Finite volume studies, including also scaling properties, are
under way.


\begin{thebibliography}{9}

\bibitem{chenetal} 
N.\ Mathur {\it et al.},
Phys.\ Lett.\ B {\bf 605} (2005) 137;
hep-lat/0405001.

\bibitem{sasaki}
S.\ Sasaki, Prog.\ Theor.\ Phys.\ Suppl.~{\bf 151} (2003) 143.

\bibitem{guadagnoli}
D. Guadagnoli, M. Papinutto and S. Simula,
Phys.\ Lett.\  B~{\bf 604} (2004) 74, and these proceedings.

\bibitem{variation}
C.\ Michael, Nucl.\ Phys.\ B~{\bf 259} (1985) 58;
M.\ L\"uscher and  U.\ Wolff,
Nucl.\ Phys.\ B~{\bf 339} (1990) 222.

\bibitem{broemmeletal} D. Br\"ommel {\it et al},
Phys.\ Rev.\ D~{\bf 69} (2004) 094513;
Nucl.\ Phys.\ B Proc.\ Suppl.\  {\bf 129-130} (2004) 251.

\bibitem{prd-paper}
T. Burch {\it et al.}, Phys.\  Rev.\  D {\bf 70} (2004)  054502;
see also hep-lat/0409014.

\bibitem{jacobi1}
C.R.~Allton {\it et al.}  [UKQCD Collaboration],
Phys.\ Rev.\ D~{\bf 47} (1993) 5128.

\bibitem{burch}
T.\ Burch, C.\ Gattringer and A.\ Sch\"afer, hep-lat/0408038.

\bibitem{jacobi2}
C.~Best {\it et al.}, Phys.\ Rev.\ D~{\bf 56} (1997) 2743;
C.~Alexandrou {\it et al.},
Nucl.\ Phys.\ B~{\bf 414} (1994) 815.

\bibitem{chirimp} 
C.~Gattringer, Phys.~Rev. D~{\bf 63} (2001) 114501;
C.\ Gattringer, I.\ Hip, C.\ B.\ Lang, Nucl.\ Phys.\ B~{\bf 597} (2001) 451.

\bibitem{bgr}
C.~Gattringer {\it et al.}  [BGR Collaboration],
Nucl.\ Phys.\ B {\bf 677} (2004) 3;
Nucl.\ Phys.\ B Proc.\ Suppl.\ {\bf 119} (2003) 796;
C.\ Gattringer, 
Nucl.\ Phys.\ B Proc.\ Suppl.\ {\bf 119} (2003) 122.

\bibitem{Luweact}
M.\ L{\"u}scher and P.\ Weisz, Commun.\ Math.\ Phys.\ {\bf 97} (1985) 59; 
G.\ Curci, P.\ Menotti and G.\ Paffuti, Phys.\ Lett.\ B~{\bf 130} (1983) 205. 

\bibitem{scale}
C.\ Gattringer, R.\ Hoffmann and S.\ Schaefer,
Phys.\ Rev.\ D~{\bf 65} (2002) 094503.

\end{thebibliography}
\end{document}